\documentclass[prl,twocolumn]{revtex4}
\usepackage{amsmath}
\usepackage{epsfig}
\usepackage{float}
\usepackage{dcolumn} 
\usepackage{multirow} 
\usepackage{mathtools}
\usepackage[utf8]{inputenc}
\usepackage{tikz}

\DeclareMathAlphabet{\mathpzc}{OT1}{pzc}{m}{it}
\newcommand{\be}{\begin{equation}}
\newcommand{\ee}{\end{equation}}
\newcommand{\bea}{\begin{eqnarray}}
\newcommand{\eea}{\end{eqnarray}}
\newcommand{\beas}{\begin{eqnarray*}}
\newcommand{\eeas}{\end{eqnarray*}}

\begin{document}
	\title
{Pseudo-symmetric random matrices: \\ semi-Poisson and sub-Wigner statistics }    
\author{Sachin Kumar$^1$ and Zafar Ahmed$^2$}
\affiliation{$~^1$Theoretical Physics Section,  $~^2$Nuclear Physics Division \\ Bhabha Atomic Research Centre, Mumbai 400 085, India}
\email{1: sachinv@barc.gov.in, 2:zahmed@barc.gov.in}
\date{\today}
\begin{abstract}
Real non-symmetric matrices may have either real or complex conjugate eigenvalues.
These matrices can be seen to be pseudo-symmetric as $\eta M \eta^{-1} = M^t$, where the metric $\eta$ could be secular (a constant matrix) or depending upon
the matrix elements of $M$. Here, we construct ensembles of a large number $N$ of pseudo-symmetric $n \times n$ ($n$ large) matrices using ${\cal N}$ $(n(n+1)/2 \le {\cal N} \le n^2)$  independent and identically distributed (iid) random numbers as their elements. Based on our numerical calculations, we conjecture that for these ensembles the Nearest Level Spacing Distributions (NLSDs: $p(s)$) are sub-Wigner as $p_{abc}(s)=a s e^{-bs^c} (0<c <2)$  and the distributions of their eigenvalues fit well to $D(\epsilon)= A[\mbox{tanh}\{(\epsilon+B)/C \}-\mbox{tanh}\{(\epsilon-B)/C\}]$ (exceptions also discussed). These sub-Wigner NLSD are encountered in  Anderson metal-insulator transition and topological transitions in a Josephson junction. Interestingly, $p(s)$ for $c=1$ is called semi-Poisson and we show that it  lies close to  the form $p(s)=0.59 s K_0(0.45 s^2)$ derived for the case of $2 \times 2$ pseudo-symmetric matrix where the eigenvalues are most aptly conditionally real: $E_{1,2}=a \pm \sqrt{b^2-c^2}$ which represent characteristic coalescing of eigenvalues in PT(Parity-Time)-symmetric systems.
\end{abstract}
\maketitle
\section{I. Introduction}
Wigner (1958) [1,2] surmised that the ensemble of $N$, $n \times n$ real symmetric  random matrices with ${\cal N}=n(n+1)/2$ independent and identically distributed (iid) matrix elements had the 
Nearest Level Spacing Distribution (NLSD: $p(s)$)  approximately as $p_W(s) = \frac{\pi}{2} s e^{-\pi s^2/4}$. Since a real symmetric matrix can be diagonalized by an orthogonal matrix and the Probability Distribution Function (PDF) is usually taken as Gaussian, this ensemble in Random Matrix Theory (RMT) [1,2] is called Gaussian Orthogonal Ensemble (GOE) and $p_W(s)$ is called GOE statistics. But when $nN$ number of levels are mixed and NLSD is found, it turns out to be Poisson: $p_P(s) =  e^{-s}$ [2]. Consequently, the nuclear levels [2] with same parity and angular momentum are well known to display Wigner's $p_W(s)$ and the mixed levels show Poisson $p_P(s)$ statistics.

However, a need of a spacing statistics intermediate to Poisson and Wigner has been argued for a long time [3]. In disordered systems the metallic phase is characterized by $p_W(s)$ and the insulator phase by $p_P(s)$ but near the Anderson's transition a $p(s)$ intermediate to these two statistics has been  expected [4].
For a disordered solids near Anderson transition, Aronov et al [5] have derived the asymptotic behaviour of $p(s)$ as $ ~e^{-A s^{2-\gamma}}$, eventually the NLSD is argued to be  $p_{AKL}(s)=B s ^\beta e^{-A \beta s^{2-\gamma}}$ which is sub-Wigner. The interesting works [6,7] could be seen as a direct confirmation of the sub-Wigner statistics [3,5] in metal insulator transitions.

Another simple and attractive form for intermediate statistics $p_{SP}(s)= 4 s e^{-2s}$ called semi-Poisson statistics has also been conjectured for Anderson's metal-insulator transition [8]. More interestingly, topological transitions in a Josephson junction [9] are also known to display such an intermediate NLSD. Arguments relating to the random sequences [10] are found underlying $p_{SP}(s)$. More recently spectral analysis of molecular resonances in Erbium isotopes [11] are also found lying close to semi-Poisson distribution. Nevertheless, a random matrix connection of sub-Wigner statistics is long due, the present work is an attempt in this direction.

For about eighteen years a conjecture of Bender and Boettcher [12] that a non-Hermitian complex PT-symmetric Hamiltonians may also have real discrete spectrum has brought a paradigm shift in physics. These Hamiltonians are invariant under the joint action Parity ($x \rightarrow -x$) and Time-reversal ($i \rightarrow -i$) and eigenvalues are either real or complex conjugate pairs below or above a critical value of a real parameter of the potential called Exceptional Point. Further, the old concept of pseudo-Hermiticity of a Hamiltonian  $\eta H \eta^{-1}= H^\dagger$ under a metric $\eta$ [13] has been re-enforced more rigorously and complex PT-symmetric  Hamiltonians have been found to be pseudo-Hermitian under $\eta=P$ [14], it therefore turns out that a pseudo-Hermitian  Hamiltonians represent PT-symmetric systems under a  generalized parity $\eta$.

Since non-Hermitian, complex PT-symmetric or pseudo-Hermitian Hamiltonians
are  becoming increasingly important so the question of their RMT can not be ignored.
By removing the Hermitian condition of the  $\beta$-ensemble [15] of tridiagonal 
matrices, an ensemble of non-Hermitian random matrices has been constructed [16] whose eigen-
values are all real. It is shown that they belong to the class of pseudo-Hermitian
operators, its statistical properties have been investigated [16]. Next, a numerical study of ensembles of random matrices of the type $M=AF$ [16] where $A$ is Hermitian is worth mentioning.
Very interestingly, it is shown how pseudo-Hermiticity, a necessary condition satisfied by operators of PT symmetric systems can be introduced in the three Gaussian classes of random matrix theory. The model describes transitions from real eigenvalues to a situation in which, apart from a residual number, the eigenvalues are complex conjugate. More recently it has been shown that such  ensemble of pseudo-Hermitian Gaussian matrices [17]  gives rise in a certain limit to an ensemble of anti-Hermitian matrices whose eigenvalues have properties directly related to those of the chiral ensemble of random matrices [18].

The fact that real symmetric and Hermitian random matrices make two separate classes of ensembles GOE and GUE having distinct statistical properties, motivates us to separate pesudo-symmetric real matrices from  the more general pseudo-Hermitian matrices [16-19]. In the present work, we invoke a  lesser class of matrices which are real pseudo-symmetric under a metric $\eta$ 
such that $\eta H \eta^{-1} =H^t$, these are essentially  non-symmetric ones. In fact, the chance that a given real matrix is non-symmetric is more than not. We can show that a real diagonalizable matrix is indeed pseudo-symmetric
under a metric $\eta=(DD^t)^{-1}$, where $D^{-1}HD=\Lambda$ (namely $D$ is the diagonalizing matrix). In more interesting, cases  $\eta$ may be secular (not depending on the elements of $H$) as a constant matrix. The secular (characteristic) equation of finding eigenvalues namely $|H-E I|=0$ is a real polynomial of $E$ which according to fundamental theorem of algebra will have only real or complex conjugate roots. Thus the real number of eigenvalues of  $n \times n$ such $N$ random matrices will have an interesting statistical distribution.
 
In a very interesting work by Edelman et al. [20], the number of real eigenvalues of a real Gaussian random $n \times n$ matrix is found to be $\sqrt{2n/\pi}$ when $n$ is large, more remarkably they also found the distribution $D(\epsilon)$ of real eigenvalues of $R$ as  an involved analytic function of $\epsilon$ and $n$.
In RMT, we raise the interesting question of NLSD for an ensemble of real (non-symmetric) pseudo-symmetric random matrices. Interesting expressions of $p(s)$ for several ensembles of $2 \times 2$ pseudo-Hermitian matrices have been derived [21-25]. Recently, we studied [26] $p(s)$ for two ensembles of $N$, $n \times n$ pseudo-symmetric random matrices. For just two real eigenvalues of cyclic matrices with ${\cal N} =5000 ,n=100$ iid matrix elements, we found $p(s)=\frac{2}{\pi}e^{-s^2/\pi}$ [26] irrespective of the (symmetric) PDF of the matrix elements. For non-symmetric tridiagonal matrices with ${\cal N}\sim n$ iid matrix elements which had all eigenvalues as real, we found the spacing statistics as $p_{\mu}(s)=\mu e^{-\mu s}$ [26].

Here in this paper, we construct several pseudo-symmetric matrices with $n(n+1)/2<{\cal N}<n^2$ and based on our numerical results, we find that the NLSD of  ensembles of  these matrices can be fitted well to  the sub-Wigner form
\begin{equation}
p_{abc}(s)=a ~s ~ e^{-bs^c}, \quad 0<c<2,\quad a\frac{\Gamma(2/c)}{cb^{2/c}}=1,
\end{equation} 
giving rise to intermediate statistic which is long desired [3,7]. Also it becomes  semi-Poisson $p_{SP}(s)$ [8,11] given above, when $c=1$. The distribution of eigenvalues of some of these ensembles fits well to
\begin{equation}
D(\epsilon)=A[\tanh\{(\epsilon+B)/C \}-\tanh\{(\epsilon-B)/C\}],  \epsilon=\frac{E}{E_{max}}
\end{equation}
or a slight variant of this. The parameters a,b,c; A,B,C do depend on $n$ and these results are insensitive to 
the choice of the probability distribution of the elements of random matrices. Remarkably, the involved expression [15] of $D(\epsilon)$ in terms of $n$ for real random matrices of order $n \times n$ with $n^2$ independent and identically distributed random elements is well represented by our proposed simple ansatz (2).
\section{II. Pseudo-symmetry of real matrices}
Let $R$ be a real square pseudo symmetric matrix with complex eigenvalues (in general) such that $\eta R \eta^{-1}= R^t$ with a real metric $\eta$. Let $R \Psi_m =\lambda_m \Psi_m$ (i) and $R \Psi_n = \lambda_n \Psi_n$ (ii). We can re-write (ii) as $\eta R \eta^{-1} \eta \Psi_n=\lambda_n \eta \Psi_n$ or $R^t  \eta \Psi_n = \lambda_n \eta \Psi_n$ (iii) . We take conjugate-transpose of (i) to write $\Psi_m^{*t} R^t= \lambda_m^* \Psi_m^{*t}$. Post multiplying this by $\eta \Psi_n$, we have $\Psi_1^{*t} R^t \eta \Psi_n= \lambda_m^* \Psi_m^{*t} \eta \Psi_n$ (iv). Pre-multiplying (iii) by $\Psi_m^{*t}$ we get $\Psi_m^{*t} R^t \eta \Psi_n=  \lambda_n \Psi_m^{*t} \eta \Psi_n$ (v). Subtracting(iv) and (v) we get
\begin{equation}
(\lambda_m^*-\lambda_n)  \Psi_m^{*t} \eta \Psi_n = 0.
\end{equation}
Here $m=n$ implies real eigenvalues and a non-zero  norm: $\Psi^{*t}_n \eta \Psi_n \ne 0$.
The cases $m\ne n$ imply (non-real) complex conjugate pairs of eigenvalues with orthogonal states:
$\Psi^{*t}_m \eta \Psi_n = 0$. So there will be a mixture of real and complex-conjugate eigenvalues.
Similarly, for anti-pseudo-symmetric matrices $\eta R \eta^{-1}=- R^t$, we can show that
\begin{equation}
(\lambda_m^*+\lambda_n)  \Psi_m^{*t} \eta \Psi_n = 0.
\end{equation}
When $m=n$, there are either real eigenvalue-pairs with opposite signs  and non-zero norm: $\Psi_n^{*t}\eta\Psi_n \ne 0$ or there are complex conjugate pairs when $m \ne n$ with orthogonal states as above.

Let the real matrix  $R$ be diagonalizable with complex  eigenvalues $\lambda_n$, in the following we prove
that it is pseudo-symmetric under $\eta=(DD^t)^{-1}$, where $D$ diagonalizes $R$. \\
{\bf Proof:}
\begin{eqnarray}
D^{-1} R D=\Lambda[\lambda_1,\lambda_2,\lambda_3,....] \Rightarrow D^{-1} \eta^{-1} \eta R \eta^{-1} \eta D =\Lambda \\ \nonumber
\Rightarrow D^{-1} \eta^{-1} R^t \eta D =\Lambda \Rightarrow (\eta D)^t R (D^{-1} \eta^{-1})^t= \Lambda  \hspace{.7cm}	\tikz \fill [black] (0.1,0.1) rectangle (0.2,0.3);
\end{eqnarray}
By comparing the first and the last parts in the above equation we get $\eta=(DD^t)^{-1}$ consistently.
\section{III. Various real pseudo-symmetric matrices}
\indent Here, we construct  real non-symmetric square matrices $R$ and  $Z$ ($Z_{ii}=$ a fixed random number), with $n^2$ and $n^2-n+1$ iid matrix elements, respectively. Let $n$ be even. 
Let us define $n \times n$ Pauli like block matrices $S_k$ using identity $I_{n/2 \times n/2}$ and null matrices $O_{n/2 \times n/2}$. $R$ and $Z$ can be called to be pseudo-symmetric under $(\eta=(DD^t)^{-1})$, where $D$ is their respective diagonalizing matrix.
\begin{equation}
S_1=\left (\begin{array}{cc} O & -I\\  I & O \end{array}\right), S_2= \left (\begin{array} {cc} O & I \\ I & O \end{array}\right), S_3=\left (\begin{array}{cc} I & O \\ O & -I \end{array}\right),
\end{equation}
such that $-S_1^2=S_2^2=S_3^2=I_{n \times n}$ and $S_k^{-1}=S_k^t=-S_k$.
Now let us construct $M_k=(R+R^t)S_k$ and $N_k=(R-R^t)S_k$ $(k=1,2,3)$ with ${\cal N}=n(n+1)/2$ number of iid matrix elements. One can readily check that $N_1,M_2,M_3$  are Pseudo-Symmetric (PS) under constant metrics $S_1,S_2,S_3$, respectively. The matrices $M_1,N_2,N_3$ are Anti-Pseudo-Symmetric (APS) under constant metrics $S_1,S_2,S_3$, respectively. We may also replace $R$ by non-symmetric  matrix $Z$ in constructing another set of such matrices.

Let us define a metric $K$-- a constant block matrix 
\begin{equation}
K=\left (\begin{array}{cc} kI & O\\ O &  I \end{array}\right), \quad K^t=K, \quad KA \ne AK, k \ne 1,
\end{equation}
which can distort a symmetric matrix into pseudo-symmetric matrix 
$P(k)=(R+R^t)K$, note that $K P K^{-1}= P^t$, namely, $P(k)$ are pseudo-symmetric under $K$. This
metric has also been utilized earlier [18]. Another class of pseudo-symmetric matrices is $Q(k)$ which is  distortion a symmetric matrix by a factor $k$ as
\begin{equation} 
%\vspace*{-10mm}
q_{ij}=x_m, \quad i \le j; \quad k x_m, \quad  i>j, \quad k>0
%q_{ij}=\left\{ \begin{array}{lcr}
%x_m, & & i \le j\\
%k x_m , & & i>j \\
%\end{array}
%\right., \quad 1   \le i,j\le n, \quad k>0.
%\vspace*{-10mm}
\end{equation}
Here $x_m$ ($1 \le m \le n(n+1)/2$) are $m$ iid random numbers under some PDF.
For $k<0$ the eigenvalues are only sparsely real.
We show that $P(k)$ and $Q(k)$ can aptly yield  spacing statistics which are
intermediate to $p_P(s)$ and  $p_W(s)$.

\section{IV. Ensembles of real pseudo-symmetric random matrices}
%\vspace*{-0.75mm}
The matrices constructed in section III are  Pseudo-Symmetric (PS) or Pseudo-Anti-Symmetric (PAS) which can represent PT-symmetric systems under the generalized parity $\eta$, their eigenvalues are real or complex conjugate pairs. $M_1,N_2,N_3$ being PAS, so as per Eq. (4) they have real eigenvalues as $\pm$ pairs. Similarly $N_1,M_2,M_3$ being PS can have real eigenvalues. All the eigenvalues of $P(k)$ if $k>0$ are all real (there are no complex conjugate pairs), but when $k<0$, there is a mixture of both real and complex conjugate pairs. The matrices $Q(k)$ have larger(lesser) number of eigenvalues as real when $k>0(k<0)$. In the literature [9,27], $M_1$  known as traceless even ordered Hamiltonian matrix with eigenvalues as quadruples: $\pm \lambda, \pm \lambda^*$.  $N_1$ is called even-ordered skew-Hamiltonian [9,28] matrix its eigenvalues are doubly degenerate. 

Earlier, $M_1$ has been suggested [27] to be make an interesting ensemble of random matrices. Ensembles of random matrices of both types $M_1$ and $N_1$ have been investigated [9] in terms of
probability of occurrence of real eigenvalues in a real random matrix. Here, we propose to investigate the ensembles of random matrices arising from pseudo-symmetric matrices $R,Z,M_k, N_k, P(k)$ and $Q(k)$ discussed above for their spectral distributions $p(s)$ and $D(\epsilon)$. We do so by considering 5000, matrices of dimension $100 \times 100$ where the matrix elements are ${\cal N}$ ($n(n+1)/2<{\cal N} <n^2$) iid random numbers under Gaussian distribution of mean 0 and variance 1. We also
use other (symmetric) probability distribution functions to observe the insensitivity of our results towards them.

\begin{figure}[h]
	\centering
	\includegraphics[width=8.5 cm,height=6.cm]{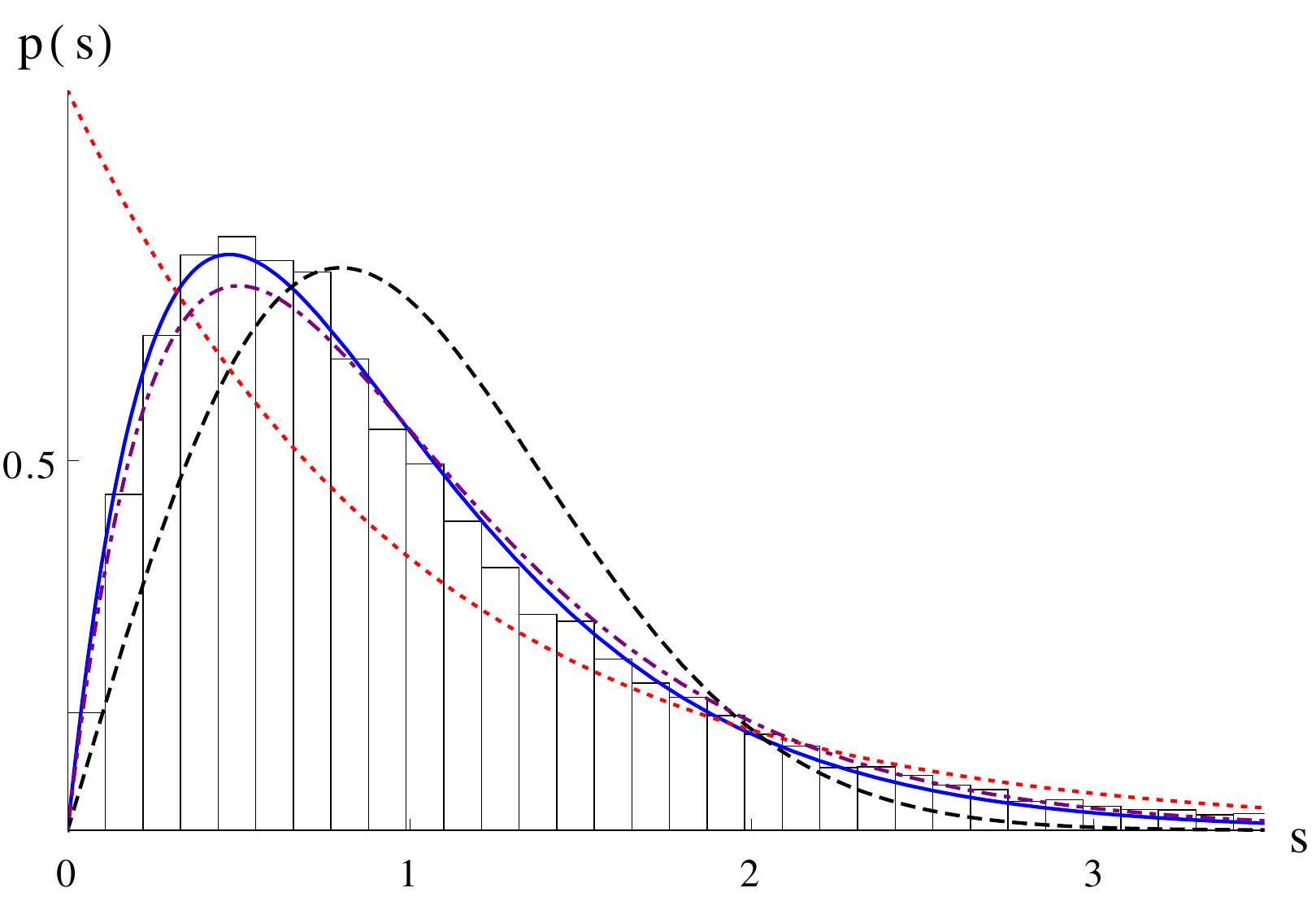}
	\caption{Semi-Poisson $p_{ab}(s)=4.50 s e^{-2.12s}$ ($c=1$ in (1)) fit to NLSD histograms  for the ensemble of 5000, $100 \times 100$ real (pseudo-symmetric) random matrices $R$. This is the typical universality also for the cases of $Z, N_k, M_1, P(-0.83), Q(0.6)$ (see Table I). The PDF of matrix elements is Gaussian but these results are insensitive to the choice of (symmetric) PDF, the parameters $a,b,c$ may change slightly for  the values of $n>100$, but the form of $p(s)=\nu^2 s e^{-\nu s}$ is robust. The dotted (red) and dashed (black) lines represent $p_P(s)$ and $p_W(s)$, respectively. The dash-dot line represents $p_{SP}(s)=4 s e^{-2s}$.}
		\end{figure}

		\begin{figure}[ht]
			\centering
			\includegraphics[width=8.5 cm,height=6.cm]{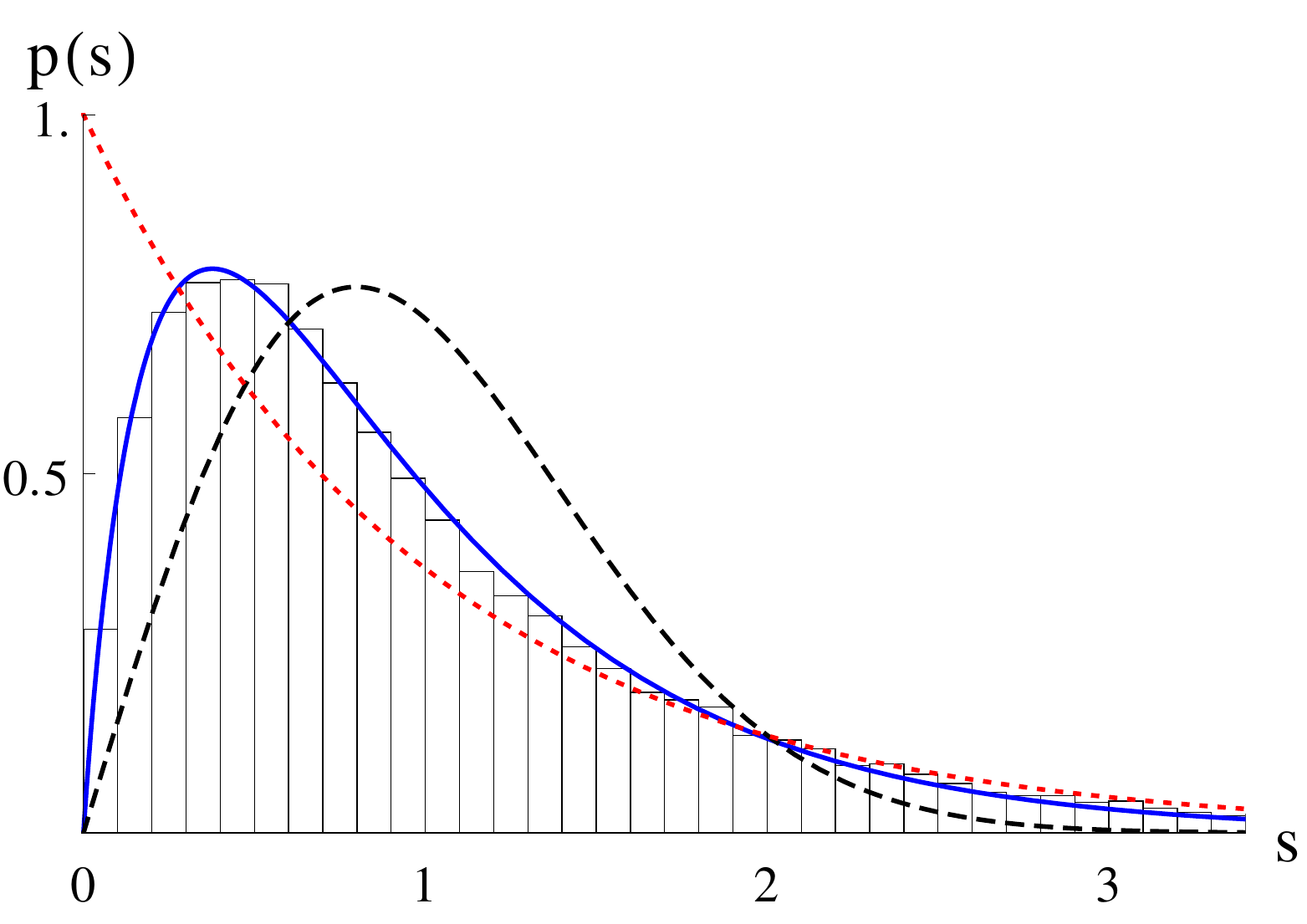}
			\caption{Sub-Wigner fit (Eq. (1), with $0<c<2$, $c \ne 1$) to the  histograms of NLSD for  ensemble of 5000, $100 \times 100$ random matrices of the type $M_2$. Here $a=7.55, b=2.75, c=0.78$. This is also a typical universality for $M_3, Q(0.30), P(\pm 0.5), P(\pm 0.9)$ (see the Table I). These results are insensitive to the PDF of matrix elements. The parameters $a,b,c$ may change slightly for $n>100$, but the form (1) is robust. The dotted (red) and dashed (black) lines represent $p_P(s)$ and $p_W(s)$, respectively.}
		\end{figure} 
			
	%\vspace{-.761cm}	
		\begin{figure}[ht]
			\centering
			\includegraphics[width=8.5 cm,height=6.cm]{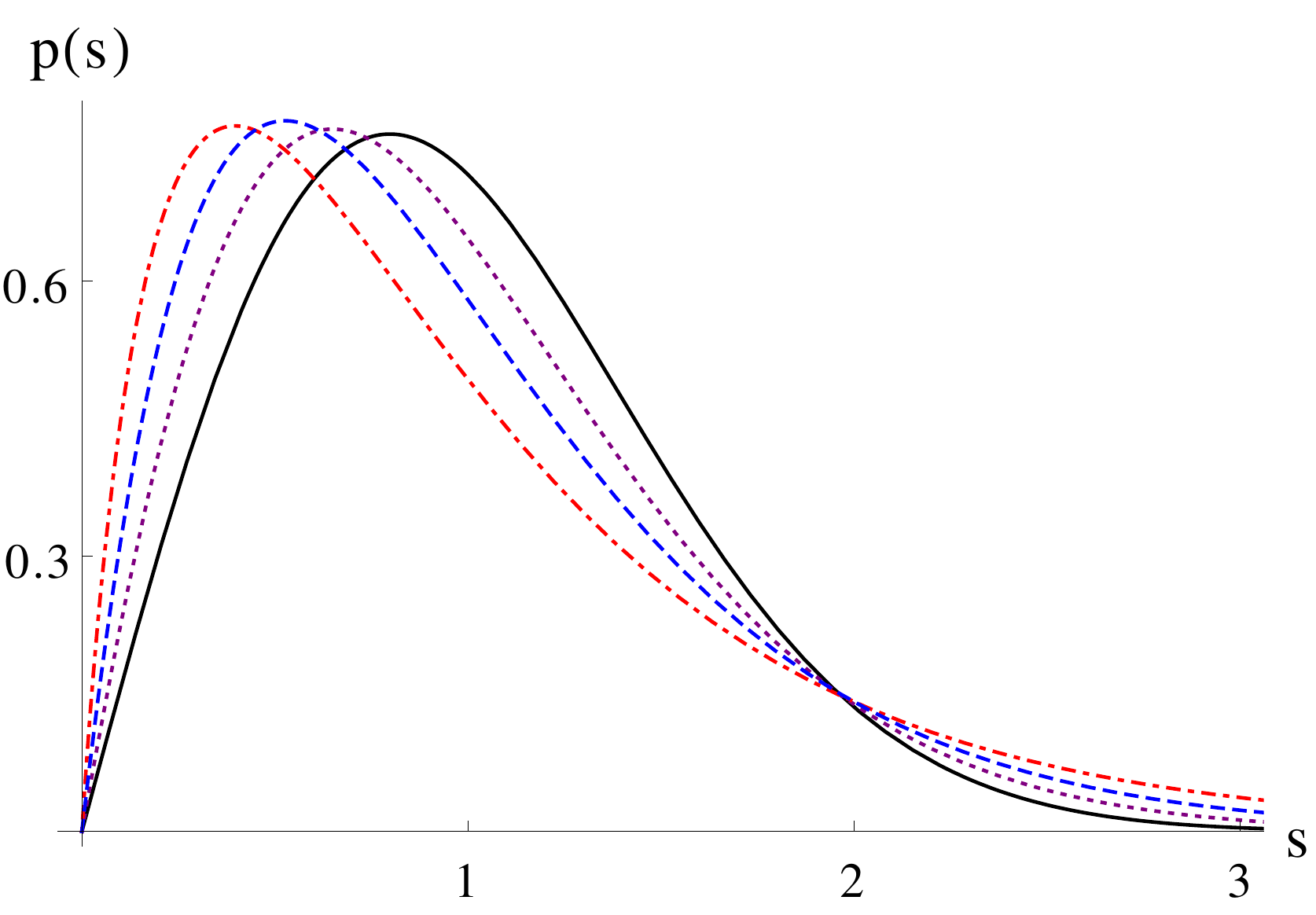}
			\caption{Fits $p_{abc}(s)= a s e^{-bs^c}$ to numerically computed histograms for the real (pseudo-symmetric) random matrix $P(k)$ sequenced from left (near y-axis) for three values of $k=-0.9$ [dot-dashed (red)], $k=-0.5$ [dashed (blue)], $k=0.9$ [dotted (purple)]  presenting intermediate $p(s)$ and solid black line displays $p_W(s)$. For the values of parameters see Table I.}
		\end{figure} 
		
	\begin{figure}[ht]
		\centering
		\includegraphics[width=8.5 cm,height=6.cm]{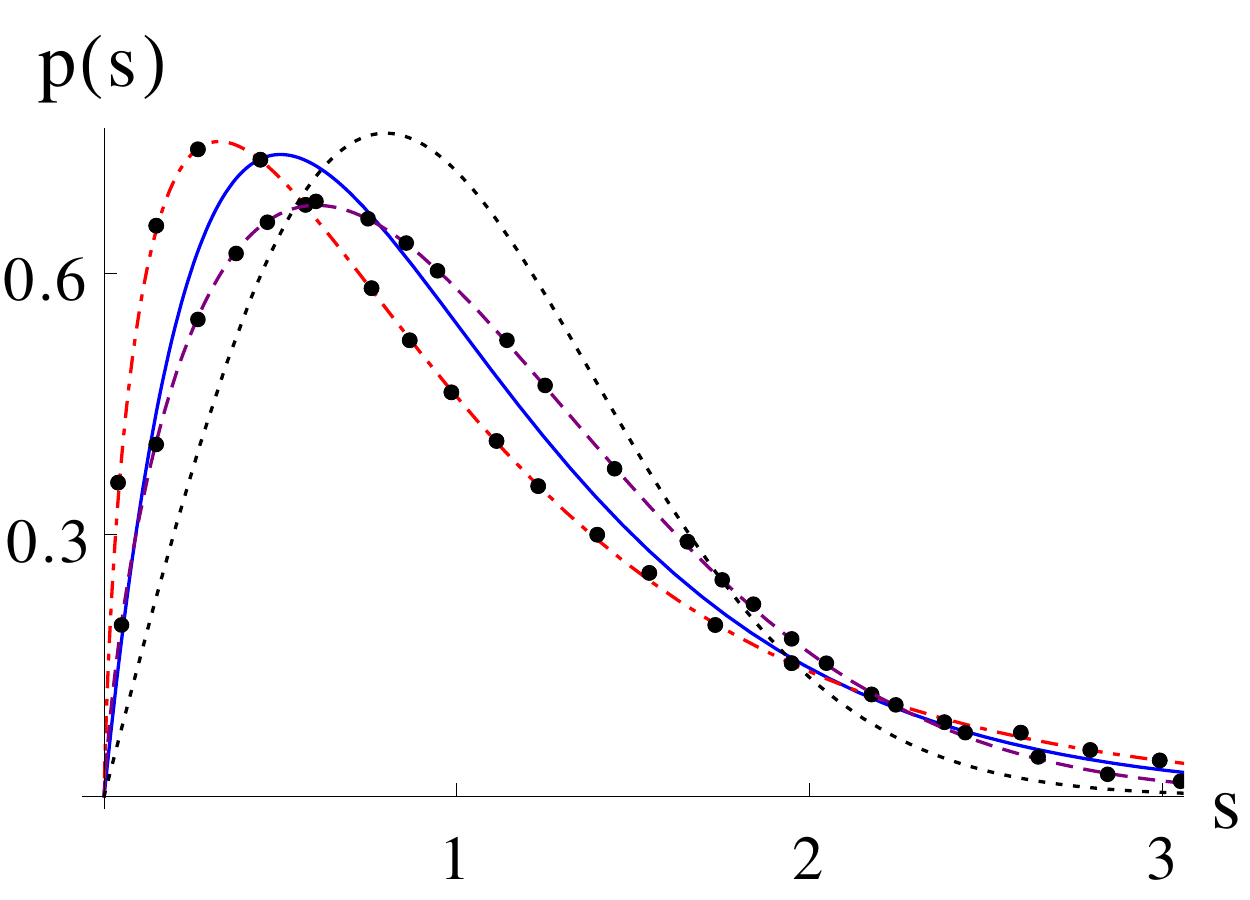}
		\caption{Spacing distributions for (9) with Gaussian [dashed (purple)] Eq. (12)) and exponential [dot-dashed (red)] Eq. (16) PDFs of matrix elements. The semi-Poisson ($4s e^{-2s}$) is given by solid (blue) line. Black dots are due to the numerically obtained histograms. All three serve as p(s) which are intermediate to Poisson and Wigner statistics [dotted (black)] line.}
	\end{figure}

\begin{figure}[ht]
	\centering
	\includegraphics[width=8.5 cm,height=6.cm]{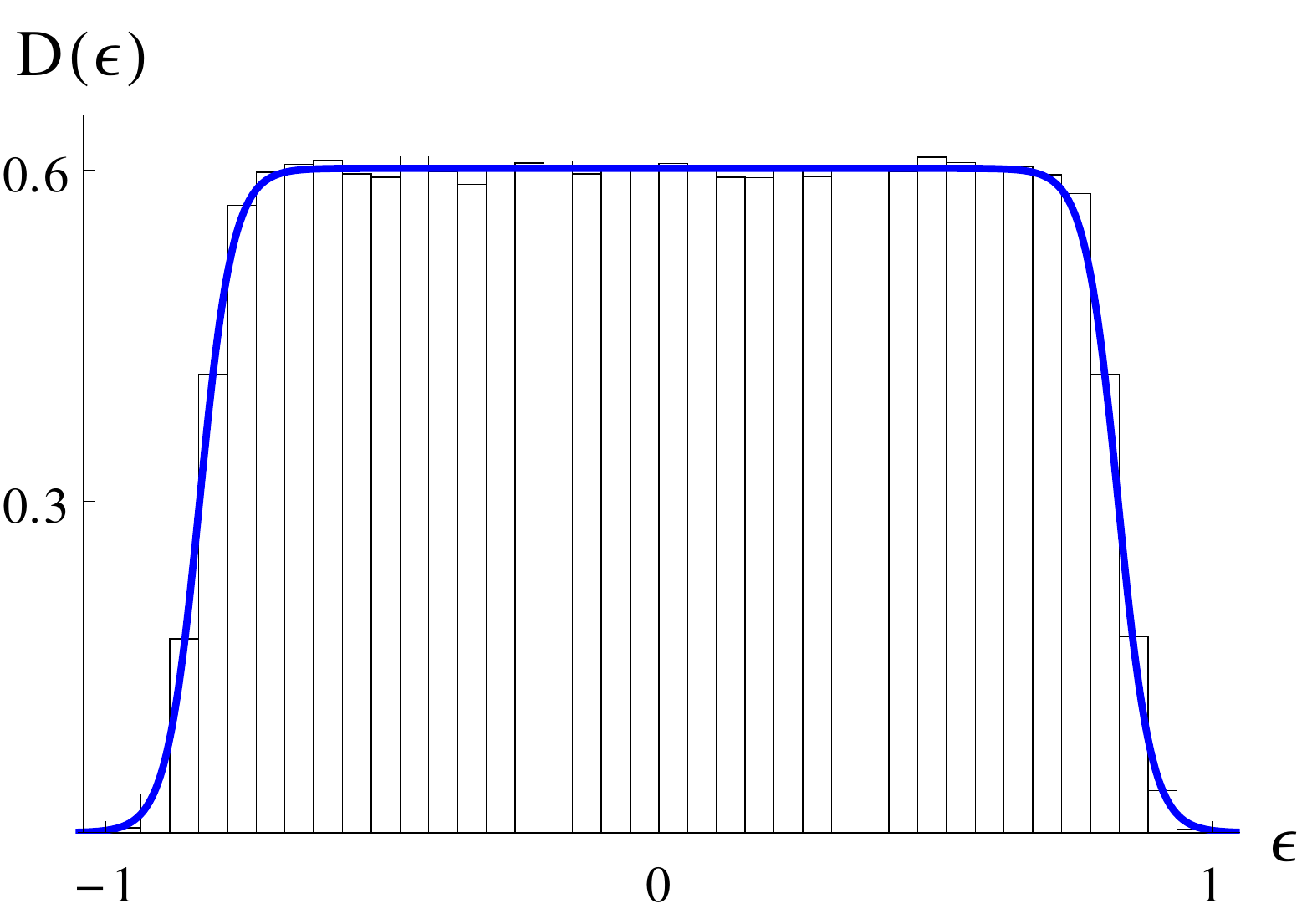}
	\caption{The distribution of eigenvalues $D(\epsilon)$ of the matrix $R$ for  5000 matrices of order $200 \times 200$. The histograms fit well to the form in
    Eq.(2), where $A=0.30, B=0.86, C=0.06$. This seems to be the universality for matrices
	mentioned in  Table 1, excepting $P(k)$ matrices.}
\end{figure} 		

	\begin{table}[]
		\centering
		\caption{The parameters of the fitted sub-Wigner form of $p_{abc}(s)$ (1) to the histograms of NLSD of  ensembles of various pseudo-symmetric matrices constructed here. The last column ensures that the area under the fitted function and histograms is almost 1 (normalized).}
		\vspace*{.5cm}
		\label{my-label}
		\begin{tabular}{|c|c|c|c|c|c|c|}
			\hline
			$Matrix$ &  $PS/PAS, \eta $ & $a$ & $b$ & $c$   & $a \frac{\Gamma{(2/c)}}{(c b^{2/c})}$   \\ \hline
			$R$    & $PS$ & \hspace{0.2cm} 4.50 \hspace{0.2cm}  &  \hspace{0.2cm}   2.12 \hspace{0.2cm}   &  1 & 1.00 \\ \hline
			$Z$   & $PS$& 4.45   &  2.10   &  1    & 1.01 \\ \hline
			$N_1$& $PS, S_1$& 4.74  & 2.19   & 1  & 0.99 \\ \hline
			$N_2$ & $PAS, S_2$&  4.75  & 2.19  &  1  & 1.00 \\ \hline
			$N_3$ & $PAS, S_3$ &  4.74  & 2.18  &  1  & 1.00 \\ \hline
			$M_1$ & $PAS, S_1$ &  4.92   & 2.23   & 1   & 0.99\\ \hline
			$Q_{0.6}$& $PS$ & 4.50 &  2.12  & \hspace{0.2cm} 1  \hspace{0.2cm}  & 1.00\\ \hline 
			$P_{-0.83}$&  $PS, K$& 4.30 &  2.09  & \hspace{0.2cm} 1  \hspace{0.2cm}  & 0.99\\ \hline 
			$M_2$ & $ PS, S_2$ &7.55   & 2.75    & 0.78     & 1.00\\ \hline
			$M_3$& $ PS, S_3$ &7.72 &  2.78  & \hspace{0.2cm} 0.78  \hspace{0.2cm}   & 1.00\\ \hline
			$Q_{0.3}$& $PS$ & 5.17 &  2.28  & \hspace{0.2cm} 0.92  \hspace{0.2cm}  & 1.02\\ \hline
			$P_{0.5}$& $PS, K$& 3.13 &  1.65  &\hspace{0.2cm} 1.25   \hspace{0.2cm}  & 1.00\\ \hline
			$P_{0.9}$& $PS, K$ & 2.35 & 1.29 &\hspace{0.2cm} 1.45 \hspace{0.2 cm} & 1.01\\ \hline
			$P_{-0.9}$& $PS, K$ &6.77 &  2.62  & \hspace{0.2cm} 0.80  \hspace{0.2cm}  & 1.01\\ \hline 
			$P_{-0.5}$& $PS, K$ &3.65&  1.84  & \hspace{0.2cm} 1.10  \hspace{0.2cm}  & 1.03\\ \hline     
			
		\end{tabular}
	\end{table}

\section{V.  $p(s)$ for a $2\times 2$ random real pseudo-symmetric matrix}
Let us now construct a  real matrix $H$  made up of three iid elements $a,b,c$ which is pseudo-symmetric under the metric $\eta$ 
\begin{equation}
H=\left (\begin{array}{cc} a-b & c\\  -c & a+b \end{array}\right), \quad \eta=\left (\begin{array}{cc} 1 & 0 \\  0 & -1 \end{array}\right),
\end{equation}
as $\eta H \eta^{-1} =H^t$. Its eigenvalues $E_{1,2}=a \pm \sqrt{b^2-c^2}$ are conditionally real $(b\ge c)$. This is the strongest characteristic of pseudo-real and PT-symmetric Hamitonians. The spacing between them is ${\cal S}=2\sqrt{b^2-c^2}$. A single parameter diagonalizing matrix $D$ for the matrix $H$ can be obtained as 
\begin{equation}
D = \frac{1}{\sqrt{\cos{2 \theta}}}\left (\begin{array}{cc} \cos{\theta} & -\sin{\theta}  \\  \sin{\theta} & -\cos{\theta} \end{array}\right), \theta= \frac{1}{2} \sin^{-1}{\frac{c}{b}}, 
\end{equation}
where, $\theta \in  (-\pi/4 ,\pi/4)$. Since $H=DED^{-1}$, so we can write $a=(E_1+E_2)/2, b=(E_2-E_1)\sec{2 \theta}/2,$ and $c=(E_1-E_2)\tan{2 \theta}/2 $. The PDF of matrix elements is $P(H)= A \exp[\mbox {-tr} (H H^{t})/(2 \sigma^2)]$ using which we derive the j.p.d.f. of eigenvalues as 
\begin{equation}
P(E_1,E_2)= A\frac{|E_1-E_2|}{2(\pi \sigma ^2)^{3/2}}  K_0 \left (\frac{(E_1-E_2)^2}{4\sigma^2} \right) e^{-\frac{(E_1+E_2)^2}{(2\sigma)^2}}.
\end{equation}  
Defining $E_1-E_2 =S$ and $E_1+E_2= U$, integrating w.r.t. $U$ from $-\infty$ to $\infty$, we get the spacing distribution $P(S)$. Further, defining $s=S/\bar{S}$, we find the normalized spacing distribution as,
\begin{equation}
p(s)=\frac{\Gamma^4(-1/4)}{32\pi^3}~s~K_0\left (\frac{2\Gamma^4(3/4)}{\pi^2} s^2\right).
\end{equation}
Alternatively, $P(S)$ can also be obtained by finding the multiple integral,
	\begin{eqnarray}
	P(S)=A \int_{-\infty}^{\infty} da \int_{-\infty}^{\infty} db\int_{-\infty}^{\infty} dc~ e^{-(a^2+b^2+c^2)/(2\sigma^2)} \\ \nonumber  \delta[S-\sqrt{b^2-c^2}], 
	\end{eqnarray}
	which gives,
	\begin{equation}
	P(S)=A' S \int_{0}^{\pi/4} e^{-S^2 \mbox{sec} 2 \theta/(2\sigma^2)}~ \mbox{sec} 2\theta~ d\theta=A' S K_0(S^2/\sigma^2).
	\end{equation}
	This in turn gives the $p(s)$ as in Eq.(12). 
	
In  Eq.(12), the coefficient of $s$ is 0.59 and that of $s^2$ is 0.45.
This spacing distribution has been encountered [22,23] by us for $2 \times 2$ (non-real) pseudo-Hermitian matrices, here we emphasize that its correspondence to the real pseudo-symmetric is more appropriate as $H$ is made up of three iid elements $a,b,c$ (like a real symmetric matrix)  The appropriate (non-real) pseudo-Hermitian $2 \times 2$ (like Hermitian matrix )  requires four iid elements $a,b,c,d$ and gives a different $p(s)$ [22,23]. We also derive $p(s)$ for (9) using PDF as uniform distribution $x \in [-1,1]$ as
\begin{equation}
p(s)=\frac{1}{2} s \sinh^{-1} \left[\frac{\sqrt{1-\pi^2s^2/36}}{s}\right], 0\le s \le 6/\pi,
\end{equation}
$p(s>6/\pi)=0$.
For exponential($e^{-|x|}$) PDF of matrix elements, we find 
\begin{equation}
p(s)=(32/9)~ s ~ \int_{4s/3}^{\infty} e^{-t}/t ~dt =(32/9)~ s~ \Gamma[0,4s/3],
\end{equation}
where $\Gamma[0,x]$ is called incomplete Gamma function.
Very interestingly,
we find that the $p(s)$ in Eqs. (12, 15, 16) show linear level repulsion near $s=0$ which is lesser with regard the Wigner surmise. This agreement is even better between  semi-Poisson distribution $p_{SP}(s)= 4s e^{-2s}$ and Eq. (12), see Fig. 4.

\section{V. Results and discussions}
Curiously among several non-symmetric matrices constructed above $Q(k)$ for $k<0$ gives real eigenvalues sparingly the probability of real eigenvalue is negligibly small. So we do not study the spectral properties 
of $Q(k<0)$. For all their matrices we take $5000$ matrices of order $100 \times 100$  ($N=5000,n=100$). We choose Gaussian PDF with mean 0 and variance 1 for matrix elements. We make histograms for the NLSD of these ensembles. The real eigenvalues of $N_1,M_2,M_3$ are doubly degenerate and those of $M_1,N_2,N_3$ occurs in $\pm$ pairs, yet their $p(s)$ is sub-Wigner (1) (see Table I). In the sub-Wigner statistics (1) there occurs the unique case of $c=1$ where $p(s)$ becomes semi-Poisson: $p_{\nu}=\nu^2 s e^{-\nu s}$, another intermediate statistics which is long desired in metal-insulator transition and in Josephson junctions.
Fig. 1, represents the $p(s)$ for the ensembles pseudo-symmetric matrices arising from $R,Z,N_1,N_2,N_3, M_1,Q(0.6)$. Here, we confirm the  finding in Ref.[9] where $p(s)$ has been found to be semi-Poisson for $M_1$ and $M_1$ is called Hamiltonian matrix [27].

When $c\ne1$, see the Fig. 2 for the ensembles of pseudo-symmetric matrices due to $M_2, M_3, Q(0.3), P(\pm 0.5), P(\pm 0.9)$. We have checked that these results are hardly sensitive to the change of PDF of matrix elements, we have used uniform and symmetric exponential PDFs as well. The parameters $a,b,c$ do depend upon $n$ the order of the matrix but the forms of $p(s)$ are robust. In these Figs. 1 and 2, $p_P(s)$ and $p_W(s)$ are plotted by dotted and dashed lines to see that the solid line fitting the histograms is (1) which lies intermediate to them for $s<1$. In Fig. 3, we demonstrate the occurrence of intermediate statistics due to $P(k)$ for $k=$ (see Table I), the solid line presents $p_W(s)$. The matrices $P(k>0)$ have ${\cal N}=n(n+1)/2$ as iid element and have all eigenvalues as real yet their $p(s)$ is not Wigner type, they are sub-Wigner this can be attributed to pseudo-symmetry. To associate with pseudo-symmetry. $P(k<0)$ which do not have all eigenvalues as real but their $p(s)$ is sub-Wigner (see Table I). Similar results as in Fig. 3 are found for ensembles of matrices matrices $Q(k>0$).

Importantly, number ${\cal N}$ of iid matrix elements considered by us in Table I lie in $(n(n+1)/2, n^2)$ and it is crucial for the observance of sub-Wigner statistics. We have also studied $p(s)$  arising from the ensembles random matrices: ${\cal C}S_k , {\cal T}S_k$ and $\Theta S_k$, where ${\cal C}, {\cal T}$, and $\Theta$ are real symmetric matrices [26] (cyclic, tri-diagonal and Toeplitz, respectively). These ensembles give $p(s)$ as Poisson or even sub-Poisson ($\sim a e^{-bs^c})$ as ${\cal N}$ in these cases is much less $\sim n$.

Though the random matrix connection of long awaited [3-11] sub-Wigner statistics has come up very well in the present work, however unfortunately we do not have the analytic results in the limit $n \rightarrow \infty$. In this regard our analytic results (12,15,16) of the $2 \times 2$ pseudo-symmetric matrix $H$ (9) under three probability distribution functions  are encouraging as all of them give rise to sub-Wigner statistics (see Fig. 4). Dots there denote the histograms obtained numerically. Since this matrix has eigenvalues which are conditionally real as it happens in the case of Parity-Time-symmetric systems [21-25], here is a strong indication between sub-Winger statistics and PT-symmetry.

The distribution of eigenvalues for most of these ensembles (Table I) of pseudo-symmetric random matrices fit well to our proposed simple empirical form (2). The random matrices $P(k)$ present exceptions by giving rise to variety of compact support profiles of $D(\epsilon)$. Also $M_1/(N_2,N_3)$  give rise to a sharp fall/rise in number of eigenvalues near $\epsilon=0$. These variations can be fitted well if in (2) we replace $A$ by $A'=A(1 \pm \alpha e^{-\beta |\epsilon|})$, where $\alpha<<1$ and $\beta>>1$. For $M_1$ this feature (fall) has also been observed in Ref. [9]. The correspondence of the proposed profile of $D(\epsilon)$ with the rigorous and involved form obtained by Edelman et al [20] for real random non-symmetric $n \times n$ matrix with ${\cal N}=n^2$ iid matrix elements is remarkable.
\vspace*{-7.mm}
\section{VI. Conclusion}
We conclude that ensembles of non-symmetric random matrices give rise to an NLSD which is sub-Wigner namely intermediate to Poisson and Wigner. This is irrespective of dimension $n (n=2, n>>2)$ of the matrices and the probability distribution of matrix elements as long as the matrix has ${\cal N}\in (n(n+1)/2,n^2)$ independent and identically distributed elements. Sub-Wigner NLSD characteristically allows more number of levels to have small spacings hence there is a lesser repulsion among eigenvalues. On the other hand for large values of $s$, the sub-Wigner distribution falls off less rapidly than a Gaussian or even an exponential.
Nevertheless, sub-Wigner distributions have been encountered in metal-insulator Anderson transitions and
Josephson junctions and this is the satisfying aspect of the present work. The  frequent occurrence  of the empirical function $D(\epsilon)$ proposed by us in Eq. (2) is another striking feature of the present work.

We have seen non-symmetric matrices as pseudo-symmetric, this has helped us to construct a variety of non-symmetric matrices rather systematically to study the abundance of sub-Wigner NLSD. Intriguingly, these matrices have variety of real discrete spectra. The spectrum could be both  fully or partially real. The real eigenvalues may be doubly degenerate. The real eigenvalues may essentially be pairs of $\pm$ values or all different. The non-symmetric matrices $Q(k)$ (8) for $k<0$ have real eigenvalues sparsely only.

Ensembles of a lesser class of random matrices called real pseudo-symmetric (having ${\cal N}\in (n(n+1)/2,n^2)$ number of iids) discussed here fill an obvious gap in Random Matrix Theory. We hope that our work which could only throw light on NLSD and distribution of eigenvalues of such ensembles marks a beginning in this regard. Unfortunately, we do not have analytic results for $n\rightarrow \infty$. We hope that the numerical results presented here will inspire the more formidable analytic results.

\end{document}